\documentclass[prb,twocolumn,amsmath,amssymb]{revtex4}

\usepackage{graphicx} 
\newcommand{\taud}{\tau_{\text{dec}}}
\newcommand{\tsim}{t_{\text{sim}}}

\begin{document}

\title{Demonstrated convergence of the equilibrium ensemble for a fast
    united-residue protein model}
\author{F.\ Marty Ytreberg\footnote{E-mail: ytreberg@uidaho.edu}}
\affiliation{Department of Physics,
    University of Idaho, Moscow, ID 83844-0903}
\author{Svetlana Kh.\ Aroutiounian\footnote{The first two authors contributed
    equally}}
\affiliation{Department of Physics, Dillard University,
2601 Gentilly Blvd., New Orleans, LA 70122}
\author{Daniel M.\ Zuckerman\footnote{E-mail: dmz@ccbb.pitt.edu}}
\affiliation{Department of Computational Biology,
    University of Pittsburgh, 3064 BST-3, Pittsburgh, PA 15213}
\date{\today}

\begin{abstract}
Due to the time-scale limitations of all-atom simulation of proteins,
there has been substantial interest in coarse-grained approaches.  
Some methods, like ``Resolution Exchange,''
[E.\ Lyman \emph{et al.}, Phys.\ Rev.\ Lett.\ {\bf 96}, 028105 (2006)]
can accelerate canonical
all-atom sampling, but require properly distributed coarse ensembles.  
We therefore demonstrate that full sampling can indeed be achieved in a
sufficiently simplified protein model, as verified by a recently
developed convergence analysis.  
The model accounts for protein backbone geometry in that rigid
peptide planes rotate according to atomistically defined dihedral
angles, but there are only two degrees of freedom
($\phi$ and $\psi$ dihedrals) per residue.  
Our convergence analysis indicates that small proteins
(up to 89 residues in our tests) can be simulated for more than
50 ``structural decorrelation times'' in less than a week on
a single processor.  
We show that the fluctuation behavior is reasonable,
as well as discussing applications, limitations, and extensions of the model.
\end{abstract}

\maketitle

\section{Introduction}
How simplified must a molecular model of a protein be for
it to allow full canonical sampling?  
This question may be important to the solution of the protein
sampling problem---the generation of protein structures properly
distributed according to statistical mechanics---because of the
well-known inadequacy of all-atom simulations, which
are limited to sub-microsecond timescales.  
Even small peptides have proven slow to reach convergence
\cite{lyman-converge}.  
Sophisticated atomistic methods, moreover, which often employ
elevated temperatures \cite{swendsen-repx,nemoto,hansmann,okamoto,garcia-repx},
have yet to show they can overcome the remaining gap in
timescales \cite{zuckerman-barriers}---which is
generally considered to be several orders of magnitude.
On the other hand, because of the drastically reduced numbers
of degrees of freedom and smoother landscapes,
coarse-grained models
(e.g., Refs.\ \onlinecite{levitt-nature,go,scheraga75,kuntz-coarse,
miyazawa,skolnick,wolynes,dill,thirumalai,
friesner,jernigan-bahar,karplus97,scheraga97a,scheraga97b,clementi-pnas,hall,
shakhnovich,voth-forcematching,zuckerman-cam})
may have the potential to aid the ultimate solution
to the sampling problem, particularly in light of recently developed
algorithms like ``Resolution Exchange''
\cite{lyman-resx,lyman-resx2}
and related methods \cite{luo-coarse,vangunsteren-resx,voth-resx}.

Although the Resolution Exchange approach, in principle, can produce
properly distributed atomistic ensembles of protein configurations,
it requires full sampling at the coarse-grained
level \cite{lyman-resx,lyman-resx2}.
While the potential for such full sampling has been suggested by
some studies of folding and conformational change
(e.g., Refs.\ \onlinecite{clementi-jmb,zuckerman-cam}),
convergence has yet to be carefully quantified in equilibrium
sampling of folded proteins.
How much coarse-graining really is necessary?
What is the precise computational cost of different approaches?
This report begins to answer these questions by studying
a united-residue model with realistic backbone geometry.

We will require a quantitative method for assessing sampling.  
A number of approaches have been suggested
\cite{brooks-converge,thirumalai-converge,vangunsteren-converge,
pande-converge,lyman-converge},
but we rely on a recently proposed statistical approach which
directly probes the fundamental configuration-space distribution
\cite{lyman-converge,lyman-converge2}.  
The method does not require knowledge of important
configurational states or any parameter fitting.  
In essence, the approach attempts to answer the most
fundamental statistical question,
``What is the minimum time interval between snapshots so
that a set of structures will behave as if each member
were drawn independently from the configuration-space
distribution exhibited by the full trajectory?''  
This interval is termed the structural decorrelation time $\taud$,
and the goal is to generate simulations of length $\tsim \gg \taud$.

In this report, we demonstrate the convergence
of the equilibrium ensemble for several proteins using a fast,
united-residue model employing rigid peptide planes.  
The relative motion of the planes is determined by the
\emph{atomistic} geometry embodied in the $\phi$ and $\psi$
dihedral-angle rotations, as explained below.  
We believe such realistic backbone geometry will
be necessary for success in Resolution Exchange studies.  
The use of geometric look-up tables enables the rapid use of
only two degrees of freedom per residue ($\phi$ and $\psi$),
and one interaction site at the alpha-carbon.  
The simulations are therefore extremely fast.  
G{\=o} interactions stabilize the native state
while permitting substantial fluctuations in the overall backbone.  

After the model and the simulation approach are explained,
the fluctuations are compared with experimental data
from X-ray temperature factors and the diversity of NMR structure sets.
The simulations are then analyzed for convergence and timing.

\begin{figure}
    \begin{center}
	\includegraphics[width=5.7cm,clip]{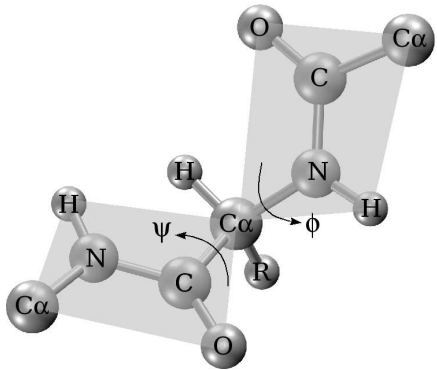}
    \end{center}
    \caption{\label{fig-rigid}
	The rigid peptide plane model used this study.
	Note that, in the coarse-grained simulations,
	only alpha-carbons are represented,
	and the only degrees of freedom are $\phi$ and $\psi$.
	Other atoms are shown in the figure only to clarify 
        the geometry and our
	assumption of rigid peptide planes.
    }
\end{figure}

\section{Coarse-grained model}
The coarse-grained model used for this study
was chosen to meet several criteria:
(i) the fewest number of degrees of freedom per residue;
(ii) the ability to utilize lookup tables for enhanced simulation speed;
(iii) the stability of the native state along with the potential
for substantial non-native fluctuations; and,
(iv) the ability to allow the addition of
chemical detail, as simply as possible.
Thus, we chose a rigid peptide plane model with G{\=o}
interactions \cite{go,go2,go-pnas}
and sterics based on alpha-carbon interaction sites as shown
in Fig.\ \ref{fig-rigid}. 
The use of such a simple model, we emphasize, is consistent
with our goal of understanding both the potential, and the
limitations of coarse models for statistically valid sampling.
Once we have understood the costs associated with the present model,
we can design more realistic models, as discussed below.
In other words, we made no attempt to design the most
chemically realistic coarse-grained model,
although we believe the use of atomistic peptide geometry
is an improvement over a coarse model we considered
previously \cite{zuckerman-cam}. 

The rigid peptide planes allows the use of only two degrees
of freedom per residue, arguably the fewest that one would consider
in such a model.
Indeed, this is fewer than in a freely rotating chain,
although admittedly our model requires somewhat more
complex simulation moves, described below.

G{\=o} interactions were used because they simultaneously
stabilize the native
state of the protein and also permit reasonable equilibrium
fluctuations, as was shown in an earlier study \cite{zuckerman-cam}.
Given our interest in native-state fluctuations and the lack of a
\emph{universal} coarse-grained model capable of stabilizing the
native state for \emph{any} protein, G{\=o} interactions are a
natural choice for enforcing stability.
Further, beyond the reasonable ``local'' fluctuations shown below,
the model also exhibits partial unfolding events which are
expected both theoretically and
experimentally \cite{falke,englander,kern}.

Because we see the present model as only a first step in the
development of better models, it is important that it easily
allows for the addition of
chemical detail, such as Ramachandran propensities which
require only the dihedral angles we use explicitly \cite{richardson}.
Furthermore, with a rigid peptide plane, the locations
of all backbone atoms---and the beta carbon---are known implicitly.
Thus hydrogen-bonding and hydrophobic interactions \cite{dill} can
be included in the model with little effort.
In other words, the ``extendibility'' of the
present simple model was a significant factor in its design.

\subsection{Potential energy of model system}
The total potential used in the model is given by
\begin{equation}
    U = U^{\rm nat} + U^{\rm non},
    \label{eq-u}
\end{equation}
where $U^{\rm nat}$ is the total energy for native contacts,
and $U^{\rm non}$ is the total energy for non-native contacts.

For the G{\=o} interactions, all residues that are separated by a distance
\emph{less} than $R_{\rm cut}$ in the experimental structure are
given native interaction energies defined by a square well:
\begin{align}
    U^{\rm nat} &= \sum_{ \{i<j\} }^{\rm native} u^{\rm nat}(r_{ij}),
	\nonumber \\
    u^{\rm nat}(r_{ij}) &= \left\{
    \begin{array}{l}
	\infty \;\; {\rm if} \;\; r_{ij} < r_{ij}^{\rm nat}(1-\delta)\\
	-\epsilon \;\; {\rm if} \;\;
	    r_{ij}^{\rm nat}(1-\delta) \leq r_{ij}
	    < r_{ij}^{\rm nat}(1+\delta)\\
	0 \;\; {\rm otherwise}
    \end{array}
    \right.,
    \label{eq-un}
\end{align}
where $r_{ij}$ is the $C_\alpha-C_\alpha$
distance between residue $i$ and $j$, $r_{ij}^{\rm nat}$
is the the distance between the residues in the experimental structure,
$\epsilon$ determines the energy scale of the native
G{\=o} attraction,
and $\delta$ is a parameter to choose the width of the well.
All residues that are separated by \emph{more} than $R_{\rm cut}$ in the
experimental structure are
given non-native interaction energies defined by
\begin{align}
    U^{\rm non} &= \sum_{ \{i,j\} }^{\rm non-native} u^{\rm non}(r_{ij}),
	\nonumber \\
    u^{\rm non}(r_{ij}) &= \left\{
    \begin{array}{l}
	\infty \;\; {\rm if} \;\; r_{ij} < (\rho_i+\rho_j)(1-\delta)\\
	+h\epsilon \;\; {\rm if} \;\; (\rho_i+\rho_j)(1-\delta)
	    \leq r_{ij} < R_{\rm cut}\\
	0 \;\; {\rm otherwise}
    \end{array}
    \right.,
    \label{eq-unn}
\end{align}
where $\rho_i$ is the hard-core radius of residue $i$ defined as half the
$C_\alpha$ distance to the nearest non-covalently-bonded residue,
and $h$ determines the strength of the repulsive interaction.

For this study, parameters were chosen to be similar to those
in Ref.\ \onlinecite{zuckerman-cam}, i.e.,
$\epsilon=1.0$, $h=0.3$, $\delta=0.1$, and $R_\text{cut}=8.0$ \AA.

\subsection{Monte Carlo simulation}
The protein fluctuations were generated using
Metropolis Monte Carlo \cite{metropolis}.
Trial configurations were generated by adding a random Gaussian
deviate to the values of three sequential pairs of backbone torsions
(three $\phi$ and three $\psi$ angles).
We found that changing six sequential backbone
torsions maximizes the rate of convergence of the equilibrium ensemble
(data not shown).
The energy of the trial configuration was
then determined using Eq.\ (\ref{eq-u}), and the conformation
was accepted with probability $\min (1,e^{-\Delta U/k_BT} )$,
where $\Delta U$ is the total change in potential energy of the system.
The width of the Gaussian distribution for generating random deviates
was chosen such that the acceptance
ratio was about 40\% for all simulations.
The choice of temperature is discussed below.

\subsection{Use of lookup tables}
The speed of the coarse-grained simulation was enhanced by using
lookup tables to avoid unnecessary computation.
In general, utilizing lookup tables increases memory
usage while decreasing the number of computations.
Since memory is inexpensive and can be expanded easily,
utilizing as much memory as possible can be an effective
technique for increasing the speed of simulations.

In our model there are only two degrees of freedom per residue
($\phi,\psi$), but $C_\alpha$ distances $r_{ij}$ must be
computed to determine native and non-native interaction energies
given by Eqs. (\ref{eq-un}) and (\ref{eq-unn}).
All peptide planes are considered to possess ideal,
rigid geometry as determined by energy minimization of
the all-atom OPLS forcefield \cite{oplsaa}
using the {\sc tinker} simulation package \cite{tinker}.

Given a sequence of three residues (alpha carbons),
we employed a lookup table to provide the Cartesian coordinates
of the third residue---starting from the N-terminus---and
its normal vector as a function of
$\phi$ and $\psi$; see Fig.\ \ref{fig-rigid}.
The table values assume that the first residue is at the origin
and the second residue is located on the z-axis. Once the coordinates
for the third residue were determined via the lookup table, the fourth
residue position was determined using the lookup table in conjunction with
a coordinate rotation and shift. Continuing in this fashion, coordinates
for the entire protein were determined.

The resolution of the lookup table is an important consideration, i.e.,
the number of $\phi,\psi$ values for which Cartesian coordinates are stored.
In our simulations, we tried resolutions as high as $0.1^\circ$
and as low as $1.0^\circ$, and found no difference between the results.
Thus, all simulation results presented here use lookup tables with a 
resolution of $1.0^\circ$.

\subsection{\label{sec-equil}Initial protein relaxation}
One perhaps unexpected complication of utilizing a rigid peptide plane model
is that great care must be taken to relax the protein
before simulations can be performed.
Although initial values of $\phi,\psi$ are obtained from the
X-ray or NMR structure,
there are slight deviations from planar/ideal geometry in
a real protein. These deviations, while small, can accumulate rapidly to
become very large differences in the Cartesian coordinate positions
of the residues.
Thus, the positions of residues near the beginning of the protein
will be nearly correct, while the residues near the end of the protein
will likely have large errors---compared to the experimental structure being
modeled---which can create severe steric clashes or even incorrect
protein topology.
The severity of these ``errors'' necessitates the use of a relaxation
procedure to generate a suitable starting structure---i.e., a set of $\phi$
and $\psi$ angles which, with our ideal-geometry peptide planes,
lead to a topologically reasonable and relatively clash-free structure.

Before we detail our relaxation procedure, we note that the need for this
additional calculation is an artifact of the simplicity of
our model which can be overcome.
With the use lookup tables, in fact,
it is possible to include \emph{flexible} peptide planes
without significantly increasing the computational
cost of the model. 
Such an approach, which does not require initial relaxation,
is currently under investigation with promising preliminary results
(data not shown).

The relaxation procedure employed in the present study first
uses the $\phi,\psi$ values directly obtained from
the experimental structure.
These dihedrals provide the initial (problematic)
structure for a coarse-grained simulation.
Due to the deviations from planarity described above,
the root means-square deviation (RMSD)
between the initial structure we create
and the experimental structure tends to be large ($\sim 10$ \AA\
was not uncommon for the proteins in this study).
To increase the number of native contacts and reduce the number
of steric clashes, we next performed what we term ``RMSD Monte
Carlo'' to relax the protein to a low RMSD structure. 
Trial moves for RMSD Monte Carlo were created as described above, but accepted
with probability $\min (1,e^{-\Delta(\text{RMSD})/k_BT_\text{RMSD}} )$,
where $k_BT_\text{RMSD} = 10^{-7}$ was chosen so that moves to a higher
RMSD were rare.
In other words, the energy function itself was not used in this initial phase.

Since residues near the beginning of the protein have less
error in the starting structure than residues near the end, we used
RMSD Monte Carlo in segments. The first twenty residues were relaxed
until the RMSD was constant within a tolerance of 0.0001 \AA, followed
by the first forty, then the first sixty and so on until the RMSD of the
entire protein was relaxed. The RMSD Monte Carlo simulation
typically brought the RMSD of the simulated structure to less than
0.5 \AA, however, there were generally still steric clashes,
and some native contacts were still not present.

The final stage of relaxation was to do regular (i.e., using energy)
Metropolis Monte Carlo simulation, with a very low temperature.

Relaxation was performed until four criteria were met:
(i) the number of native contacts in the relaxed structure
was equal to that in the NMR or X-ray structure;
(ii) no steric clashes were present;
(iii) no non-native contacts were present,
i.e., $U^{\rm non} = 0$ in Eq.\ (\ref{eq-unn}), and;
(iv) the RMSD was less than 1.0 \AA.
When these criteria were
met the structure was saved and used
as the starting configuration in all future simulations of the protein.

\begin{figure*}
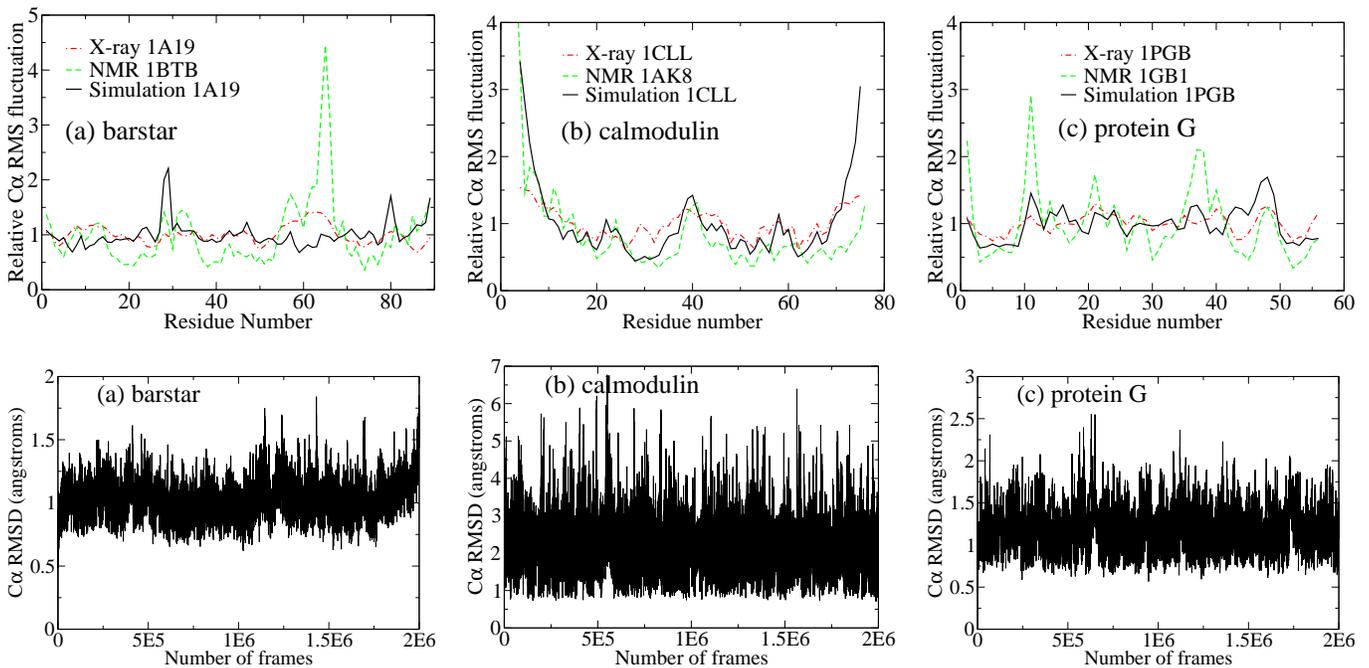

    \begin{center}
	\includegraphics[width=5.7cm,clip]{rmsf-bar.eps}
	\hfill
	\includegraphics[width=5.7cm,clip]{rmsf-cam.eps}
	\hfill
	\includegraphics[width=5.7cm,clip]{rmsf-pg.eps}
    \end{center}
    \begin{center}
	\includegraphics[width=5.7cm,clip]{rmsd-1a19.eps}
	\hfill
	\includegraphics[width=5.7cm,clip]{rmsd-1cll.eps}
	\hfill
	\includegraphics[width=5.7cm,clip]{rmsd-1pgb.eps}
    \end{center}
    \caption{\label{fig-rmsf}
	(Color online)
	Relative alpha-carbon root mean square fluctuations for three
	different proteins: (a) barstar, (b) calmodulin, and (c) protein G.
	Each plot shows results for the
	X-ray structure (dot-dash), the NMR ensemble (dash),
	and the coarse-grained simulation (solid).
	X-ray results were given by $\sqrt{3B/8\pi^2}$, where
	$B$ is the temperature factor given in the PDB entry.
	NMR and simulation data were generated using the
	g\_rmsf program in the {\sc gromacs} molecular simulation
	package \cite{gromacs}; each ensemble was aligned to the
	first structure in the corresponding trajectory.
	For each coarse-grained simulation, $2\times 10^9$ Monte
	Carlo steps were performed with snapshots saved every
	1000 steps, and the potential energy
	\eqref{eq-u} was set up using the X-ray structure.
        Panels (d) - (f) show the corresponding whole-structure
	fluctuations as indicated by the RMSD from the experimental structures.
    }
\end{figure*}

\section{Results and Discussion}
Using the coarse-grained protein model described above, we
generated and tested equilibrium ensembles for three proteins:
barstar (PDB entry 1A19, residues 1-89),
the N-terminal domain of calmodulin (PDB entry 1CLL, residues 4-75), and
the binding domain of protein G (PDB entry 1PGB, residues 1-56)

For each protein, the initial simulation structure was generated,
followed by RMSD and energy relaxation, as described in
Sec.\ \ref{sec-equil}. Then, production runs of
$2 \times 10^9$ Monte Carlo moves were performed with snapshots
saved every 1000 moves, generating an equilibrium ensemble
with $2 \times 10^6$ frames.

In an attempt to obtain consistent results for the three proteins, 
we chose the temperature of the simulation, $k_BT$, to
be slightly below the unfolding temperature of the protein. The unfolding
temperature was determined by running simulations over a broad range
of temperatures and studying the RMSD as a function of simulation
time. The temperatures used in the simulations were $k_BT=0.6$ for barstar,
$k_BT=0.4$ for calmodulin and $k_BT=0.5$ for protein G.

\subsection{Speed of simulations}
Due to the use of lookup tables for coordinate transformations,
the small number of degrees of freedom,
and utilizing simple square potentials, the equilibrium
ensembles were generated very rapidly.

Running on one Xeon 2.4 GHz processor, $2 \times 10^9$ Monte
Carlo moves with snapshots saved every 1000 steps took roughly
6 days for barstar, 4 days for calmodulin, and 3 days for protein G.
Thus, less than a week was required to obtain well-converged
(see Sec.\ \ref{sec-conv}) simulations
of these coarse-grained proteins.

\subsection{Protein fluctuations}
We first sought to determine whether fluctuations in the
coarse-grained simulation are reasonable.
Figure \ref{fig-rmsf} shows the alpha-carbon
relative root mean square fluctuation for three
different proteins.
The figures show that there is reasonable qualitative agreement
between the NMR, X-ray and simulation data.

It should be noted that, in fact, \emph{none} of the three
data sets in Figs.\ \ref{fig-rmsf}a, b and c represents the true
fluctuations in the protein---for different reasons.
The X-ray temperature factor, in addition to thermal fluctuations,
includes crystal lattice artifacts and other experimental errors \cite{northrup}.
NMR ensembles tend to be biased, perhaps severely, toward low energy structures, and
thus also do not represent equilibrium ensembles \cite{spronk}.
Finally, our simulation data is
not accurate due to the lack of chemical detail in the forcefield.

In spite of the limitations of the analysis, we conclude
from Fig.\ \ref{fig-rmsf} that
the fluctuations of the coarse-grained model are in fact
reasonable.

The bottom panels of Fig.\ \ref{fig-rmsf} show the whole-molecule
fluctuations exhibited throughout the trajectories.
In addition to the ability to sample large conformational
fluctuations---such as in the case of calmodulin and,
to a lesser degree, for protein G---the trajectories are
visibly more converged than is typically observed in atomistic
simulations, where RMSD values rarely reach a plateau value,
let alone sampling around that plateau value multiple times
as would be desirable.

\subsection{\label{sec-conv}Convergence analysis}
The primary purpose of this report is to demonstrate the convergence
of the equilibrium ensemble for a coarse-grained protein.
The details of the convergence analysis are described in
Ref.\ \onlinecite{lyman-converge2}, so we will only briefly describe
the method here.

Previously, Lyman and Zuckerman \cite{lyman-converge}
developed an approach which groups sampled conformations
into structural histogram bins, using the RMSD as a metric.
While promising, the primary limitation of the method was
the lack of a quantitative measure of the convergence.

In the method used here, convergence
was analyzed by studying the variance of the structural histogram bin
populations \cite{lyman-converge2}. 
The new approach allows a rigorous
\emph{quantitative} estimation of convergence---the structural
decorrelation time
$\taud$, given by the time between frames required for the
variance to reach an analytically computable independent-sampling value.
Intuitively, and mathematically, $\taud$ is the time interval
between snapshots for which they behave as if each frame
were drawn independently. 
If simulation times $\tsim \gg \taud$
are obtained, the equilibrium ensemble is considered converged.

Perhaps the most important feature of the convergence
analysis for our study is that the method does not require
any prior knowledge of important states.
Furthermore, there is no parameter-fitting or subjective analysis of any kind.

\begin{figure*}
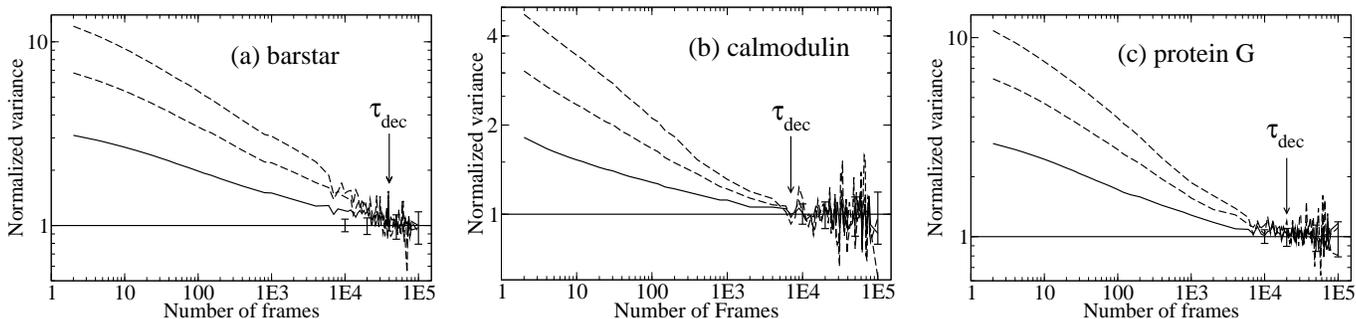

    \begin{center}
	\includegraphics[width=5.7cm,clip]{conv-bar.eps}
	\hfill
	\includegraphics[width=5.7cm,clip]{conv-cam.eps}
	\hfill
	\includegraphics[width=5.7cm,clip]{conv-pg.eps}
    \end{center}
    \caption{\label{fig-conv}
	Convergence analysis for coarse-grained simulations of
	three different proteins:
	(a) barstar, (b) calmodulin, and (c) protein G.
	Each plot shows the convergence properties for the same trajectories
	as used for Fig.\ \ref{fig-rmsf},
	analyzed using the procedure in
	Ref.\ \onlinecite{lyman-converge2}.
	The number of frames required to reach the value of one
	(the solid horizontal line) is an approximation
	of the structural decorrelation time $\taud$
	and is shown on each plot.
	The three curves on each plot are results for different
	histogram sub-sample sizes \cite{lyman-converge2}
	and demonstrates the robustness
	of the value of $\taud$.
	The plots predict that the decorrelation times are roughly
	40 000 frames for barstar, 20 000 frames for calmodulin
	and 30 000 frames for protein G.
	Note that the total number of frames generated for each protein
	during the simulation was $2\times10^6$.
	Thus, since each simulation was more than $50 \taud$
	in length, we conclude that the equilibrium ensembles
	are well-converged.
        Error bars represent 80\% confidence intervals
	in the expected fluctuations around the ideal value of one,
	based on the given trajectory length and the numerical procedure
	used to generate the solid curve.
    }
\end{figure*}

Figure \ref{fig-conv} shows the convergence properties
of the coarse-grained simulations using the same trajectories 
as in Fig.\ \ref{fig-rmsf}.
The ratio of the observed variance to the ideal variance for independent
sampling is plotted as a function of the time between the configurations
used to compute the observed variance.
When this ratio decreases to one the structural decorrelation time $\taud$
has been reached, as shown in the figure.
The analysis predicts that each simulation
is at least 50 times longer than the
structural decorrelation time.

Thus we conclude that, in less than a week of single-processor
time, the equilibrium ensembles for these three proteins are
well converged.

\section{Conclusions}
We have demonstrated the convergence of the equilibrium 
ensemble for a simple united-residue protein model.
The model assumes rigid peptide planes, with atomistically
correct geometry, and exhibits reasonable residue-level
fluctuations based the planes' geometry, G{\=o} interactions,
and sterics.

Most importantly, the results indicate \emph{quantitatively}
that carefully designed united-residue models have
the potential to fully sample protein fluctuations.
By using only two degrees of freedom per residue,
look up tables for coordinate transforms, and
simple square well potentials, we were able
to demonstrate that converged equilibrium ensembles
can be obtained in less 
than a week of single processor time.
The quantitative convergence analysis indicates that more than 50
``decorrelation times'' were simulated in each case,
indicating high-precision ensembles.
In addition to application in Resolution Exchange sampling of
all-atom models \cite{lyman-resx,lyman-resx2},
such speed opens up the long-term possibility of large-scale
simulation of many proteins.

One important practical limitation of the ideal-peptide-plane
geometry in the present model is the need to relax
the the initial structure.
Proteins larger than 100 residues are difficult to relax.
However, we have already begun investigating a flexible-plane
model incorporating lookup tables which exhibits no such
limitation and remains computationally affordable.
We will report on the flexible model in the future.

Although the intrinsic atomistic geometry of the peptide
plane was included in our model, it lacks chemical interactions.
Yet because we obtained converged ensembles in such a short
time, it is clear we can ``afford'' extensions
to the model which include realistic chemistry.
For instance, additional potential energy terms such as
Ramachandran propensities \cite{richardson},
hydrophobic interactions \cite{dill}
and hydrogen-bonding can be included at small cost.

Aside from the potential for rigorous atomistic
sampling \cite{lyman-resx,lyman-resx2,ytreberg-bbrw},
it is important to note the general usefulness of coarse-grained
models for generating \emph{ad hoc} atomistic ensembles.
Specifically, upon generating a well-sampled ensemble of coarse-grained
structures, atomic detail can be added using existing software
such as those in Refs.\ \onlinecite{sccomp,rapper}.
Once minimized and relaxed,
these (now) atomically detailed structures form
an \emph{ad hoc} ensemble which 
may be of immediate use in docking \cite{knegtel,shoichet-nature}
and homology modeling applications.
Further, in principle, such structures can be re-weighted
into the Boltzmann distribution
\cite{ytreberg-bbrw}.

In the long term, one can imagine a day when structural databases
will be based not on single (static) structures but rather
will collect ensembles---as envisioned in the authors' scheme for an
``Ensemble Protein Database''(http://www.epdb.pitt.edu/).

\begin{acknowledgments}
We thank Edward Lyman, Bin Zhang and Artem Mamonov
for helpful discussions.
Funding was provided by the National Institutes of Health
under fellowship GM073517 (to F.M.Y.),
and grants GM070987 and ES007318, 
and by the National Science Foundation grant MCB-0643456.
\end{acknowledgments}

\bibliography{/home/marty/res/tex/my}

\end{document}